\documentclass[%
 reprint,
superscriptaddress,
showpacs,preprintnumbers,
 amsmath,amssymb,
 aps,
 pra,
]{revtex4-1}

\usepackage{graphicx}
\usepackage{dcolumn}
\usepackage{bm}
\usepackage{hyperref}

\newcommand{\ie}{i.e.\ }
\newcommand{\eg}{e.g.\ }
\newcommand{\dt}{dt\,} 
\begin{document}


\title{Superradiance and collective gain in multimode optomechanics}

\author{T.~Kipf} \affiliation{Institut f\"{u}r Optik, Information und Photonik,
Universit\"{a}t Erlangen-N\"{u}rnberg, Erlangen, Germany}
\author{G.~S.~Agarwal} \affiliation{Department of Physics, Oklahoma State
University, Stillwater, OK, USA}

\date{\today}

\begin{abstract} We present a description of a strongly driven multimode
optomechanical system that shows the emergence of cooperative effects usually
known from systems of atom-light interaction. Our calculations show that under
application of a coherent pump field the system's response can be switched from
a superradiant regime to a collective gain regime by varying the frequency
detuning of the pump. In the superradiant regime, enhanced optical cooling of a
single vibrational mode is possible, whereas the collective gain regime would
potentially enable one to achieve almost thresholdless phonon laser action. The
threshold pumping power scales as $1/N$.  \end{abstract}


\maketitle \section{Introduction} Collective scattering effects based on
coherent interaction of resonant systems have been of interest since the
seminal work by Dicke in 1954 \cite{Dicke1954}.  Systems of atom-light
interaction showing superradiance have since been studied both theoretically
\cite{Rehler1971, Agarwal1974, HarocheGross1982} and experimentally
\cite{Feld1973, Scully2006, Haroche1982} for many years, but have gained
increasing attention only in the last decade owing to progress in laser cooling
of atomic clouds \cite{Baumann2010,  Kaiser2010}. These systems allow the
direct observation of cooperative scattering, but are limited in their range of
experimentally accessible parameters. Furthermore, the coherent control of
atoms within small distances, that are required for this task, is generally
difficult to achieve.  In recent years, interest has shifted to a new class of
systems of artificial atoms, such as quantum dots \cite{Scheibner2007} and
Cooper pair boxes \cite{Spiller2004}, that were found to show analogous
collective effects such as in ensembles of atoms. With the rapid advances in
the field of optomechanics, both in the optical \cite{Metzger2004, Gigan2006}
and the microwave regime \cite{Massel2012, Suh2014}, new candidates have
emerged for the study of collective behavior on the quantum level. The
on-circuit implementation of the optomechanical interaction at microwave
wavelengths \cite{Massel2012, Suh2014} hereby introduces the possibility of
coupling multiple nano-mechanical oscillators to a common cavity, thus offering
a versatile approach to studying cooperative dynamics over a wide range of
parameters.

In this paper, we present a theoretical description of a multimode
optomechanical system with regard to the emergence of collective behavior. The
flexibility of this system lies in the possibility of bringing it from a
superradiative state to a state, where collective gain can be observed, by
varying the detuning of the driving pump. We derive the collective equations
governing the dynamics of the system starting from an optomechanical
Hamiltonian description and discuss the dependence of the system on its
parameters.  We specialize in the case of $N=2$ mechanical oscillators to
provide a physical explanation for the emergence of superradiance and
collective gain and then generalize to an arbitrary number of oscillators.

\begin{figure}[htp] \includegraphics[width=0.33\textwidth]{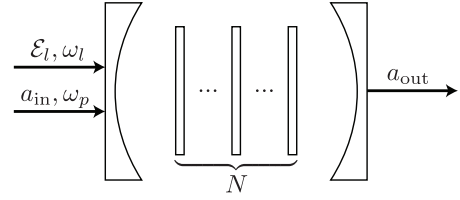}
\caption{\label{fig:setup} Schematic of the (theoretically) studied
optomechanical system in multimirror assembly, consisting of $N$ mechanical
oscillators coupled to a common cavity. The system is strongly driven with a
pump of frequency $\omega_l$ and electric field strength $\mathcal{E}_l$. The
output field $a_{\text{out}}$ is studied for an input probe field
$a_{\text{in}}$ of frequency $\omega_p$.} \end{figure}

We mention that collective effects in optomechanical systems have attracted
some attention in recent years. Shkarin et al. \cite{Shkarin2014} and Buchmann
et al. \cite{Buchmann2014} have noticed how collective effects can lead to the
coupling of two mechanical oscillators which can be used to transfer energy
from one mechanical oscillator to the other. Mumford et al. \cite{Mumford2014}
have studied the possibility of a Dicke-type phase transition in a system
involving two cavity modes and one phonon mode. In extensive studies Xuereb et
al. \cite{Xuereb2012, Xuereb2013, Xuereb2014} consider the possibility of long
range interactions in optomechanical arrays. They show extreme sensitivity of
the optomechanical interactions to the net reflection coefficient of the
dielectric array \cite{Xuereb2012, Xuereb2013}. They report \cite{Xuereb2014}
collective behavior of the array in the bad cavity limit, \ie when
optomechanical coupling $\ll$ mechanical frequency $\ll$ cavity damping. The
collective behavior that we report is in a different regime of parameters,
which is directly relevant to superconducting electromechanical systems
\cite{Massel2012, Suh2014} and graphene-based systems in superconducting
resonators \cite{Singh2014}.

Whereas we present our analysis and results with regard to the set of systems
as in \cite{Massel2012, Suh2014,Singh2014}, the extensive work of Xuereb et al.
\cite{Xuereb2012, Xuereb2013, Xuereb2014} brings out many new possibilites
which depend on the overall reflectivity of the mechanical array, making the
optomechanical coupling vary significantly from one element to the other.

This paper is structured as follows. In Section II we introduce our model,
which is based on recent experimental progresses in on-circuit implementations
of optomechanics and derive a linearized Hamiltonian description the system. In
Section III we analytically solve the system's equations for the response
function under resonant driving on the anti-Stokes sideband. We briefly discuss
the emergence of superradiant collective behavior resulting from the coupling
of the array of similar mechanical oscillators to a common reservoir. This
superradiance behavior in such systems is free from the complications arising
from the dipole-dipole interactions which can destroy superradiance in atomic
ensembles \cite{Friedberg1973}. In Section IV we extend this analysis for the
case of driving on the Stokes sideband. After deriving a criterion for stable
operation of the optomechanical system, we analyze the characteristics of the
response function. Here, our calculations suggest that the coupling mediated by
the cavity field leads to collective gain in the output field of the cavity. We
conclude in Section V.

\section{Model} Let us consider a system of $N$ independent mechanical
oscillators coupled to a common photonic cavity strongly driven with a pump of
frequency $\omega_l$ as schematically shown in Fig.\ \ref{fig:setup}.
Experimentally, such a system can be realized in an on-circuit implementation
using electro-mechanical capacitors connected to a superconducting microwave
resonator, as recently demonstrated in \cite{Massel2012, Suh2014}. Via
capacitive coupling, the mechanical oscillators modulate the resonance
frequency of the common microwave cavity. As opposed to optical systems, the
on-circuit implementation is coherently driven with microwave photons and can
in principle hold an arbitrary number of mechanical oscillators with individual
resonance frequencies. With that, these systems are a promising candidate for
the investigation of collective effects in ensembles of harmonic oscillators,
as theoretically described in \cite{Agarwal1971}.  

The optomechanical Hamiltonian of our proposed system is given by
\begin{equation} \label{eq:hamiltonian} H = \hbar \omega_c c^\dagger c
+\textstyle{\sum}_{j=1}^N [ \hbar \omega_j b_j^\dagger b_j - \hbar c^\dagger c
g_j (b_j^\dagger + b_j)] + H_l , \end{equation} where $c$ and $b_j$ are the
bosonic annihilation operators for the cavity mode and the mechanical modes,
respectively. The cavity resonance frequency is given by $\omega_c$, $\omega_j$
denotes the resonance frequency of the $j$-th mechanical oscillator and $g_j$
is the optomechanical coupling rate.

The system is strongly driven with a pump of frequency $\omega_l$ and power
$P_l$. The corresponding Hamiltonian reads \begin{equation} H_l = i \hbar
\mathcal{E}_l ( e^{-i\omega_l t}c^\dagger - e^{i\omega_l t}c), \end{equation}
with an amplitude $\mathcal{E}_l = \sqrt{2\kappa_E P_l / \hbar \omega_l }$,
where $2\kappa_E$ is the cavity linewidth associated with external coupling.
The total cavity linewidth is given by $2\kappa = 2\kappa_E + 2\kappa_I$,
whereas $\kappa_I$ accounts for all internal losses. 

In a frame rotating with the pump frequency $\omega_l$, linearized about a
steady state, the Hamiltonian reads \begin{equation} \label{eq:LinHamiltonian}
H \approx \hbar \Delta a^\dagger a + \hbar \textstyle{\sum}_{j=1}^N [\omega_j
b_j^\dagger b_j - (G_ja^\dagger + G_j^*a) (b_j^\dagger + b_j)] , \end{equation}
where we have defined the enhanced coupling rate as $G_j = \alpha g_j$ and $c
\equiv \alpha + a$, with the system's steady-state amplitude $\alpha =
\mathcal{E}_l / (\kappa + i\Delta)$. $\Delta = \omega_c - \omega_l$ is the
detuning from the cavity resonance frequency. Note that we have dropped purely
classical and small terms. Note also that $a$ is slowly varying as we are in
the rotating frame.

\section{Superradiance} In the following, we work in the resolved sideband
regime $\omega_m \gg \kappa$ and assume similiar mechanical resonators with
$\omega_j\approx \omega_m$. For cooperative effects to occur, the coupling
needs to become resonant. This can be achieved by choosing the pump frequency
$\omega_l$ such that $\Delta \approx \pm \omega_m$. With this choice, the
slowly varying intracavity field oscillates at the mean frequency of the
mechanical oscillators.

Let us first consider driving on the anti-Stokes sideband, \ie $\Delta \approx
\omega_m$. The physical process that we consider in this section corresponds to
the generation of a phonon $-\omega_l + \omega = \omega_m$ which then can
combine with a pump photon $\omega_l$ to produce an anti-Stokes photon
$\omega_l+\omega_m$. Here, the interaction terms $G_j a^\dagger b_j^\dagger$
and  $G^*_j a b_j$ become off-resonant and can be neglected in the rotating
wave approximation (RWA). 

We introduce dissipative dynamics, accounting for leakage of photons and
phonons, in form of the quantum Langevin equations for the Heisenberg operators
$a$ and $b_j$ \begin{subequations} \label{eq:quantumLangevinEquations}
\begin{align} \dot{a} &= -(\kappa + i\Delta) a + i
\textstyle{\sum}_{j=1}^NG_jb_j + f_a(t)   , \\ \dot{b}_j &=
-(\Gamma_j+i\omega_j)b_j + iG_j^* a + f_{b_j}(t)  .  \end{align}
\end{subequations} where $f_i(t)$ are the quantum Langevin forces, which
account for vacuum noise and any thermal noise entering the system. The
correlation functions associated with the quantum and thermal fluctuations are
given by \cite{Agarwal2012}: \begin{subequations} \label{eq:correlators}
\begin{align} \langle f_a^\dagger(t)f_a(t')\rangle &= 0, \\ \langle
f_a(t)f_a^\dagger(t')\rangle &= 2\kappa\delta(t-t'),\\ \langle
f_{b_j}^\dagger(t)f_{b_j}(t')\rangle &= 2\Gamma_j\bar{n}_j\delta(t-t'), \\
\langle f_{b_j}(t)f_{b_j}^\dagger(t')\rangle &=
2\Gamma_j(\bar{n}_j+1)\delta(t-t'),\\ \langle
f_{b_j}^\dagger(t)f_{b_k}(t')\rangle  &= 0 \quad (\forall j \neq k) .
\end{align} \end{subequations} Here, $\bar{n}_j$ denotes the thermal occupation
of the heat bath associated with the mechanical mode $b_j$. Note, that we have
adopted the standard Markov approximation for the correlation functions given
in Eqs.\ \eqref{eq:correlators}, \ie we have assumed delta-correlated noise
without memory.

For the case of $N=2$ degenerate mechanical oscillators with $\omega_1 =
\omega_2$, the form of the quantum Langevin equations
\eqref{eq:quantumLangevinEquations} has been studied in \cite{Jha2013}. Here,
we focus on the more general case of \textit{near-degenerate} mechanical
oscillators and study the emergence of collective behavior depending on the
detuning of the mechanical oscillator frequencies. To proceed, we set $f_a(t) =
\sqrt{2\kappa_E}a_{\text{in}}(t)$, hence assuming that a probe field
$a_{\text{in}}(t)$, strong enough compared to single photons but yet much
weaker than the pump, is applied. For simplicity, we also neglect the force
terms $f_{b_j}(t)$.

\begin{figure}[htp] \includegraphics[trim=0.5cm 0.7cm 0.4cm 0cm, clip=true,
width=0.44\textwidth]{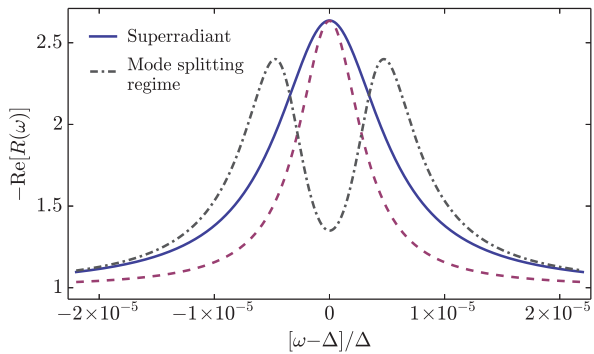} \caption{\label{fig:RedDetuned}
(Color online) Response function for $\Delta\omega = 0$ and $N=2$ (solid curve)
showing an increased linewidth compared to a system with $N=1$ (dashed red
curve, scaled and translated to facilitate comparison). The black dash-dotted
curve corresponds to $\Delta\omega = 4.5\Gamma$ and shows clearly separated
modes. Parameters were chosen as follows: $\kappa/2\pi = 1.0 \ \text{MHz}$,
$\omega_m=10\kappa$, $\Gamma=10^{-4}\kappa$, $G_1 = G_2 =
1.5\sqrt{\kappa\Gamma}$, close to typical experimental values \cite{Massel2012,
Suh2014}.} \end{figure}

We solve Eqs.\ \eqref{eq:quantumLangevinEquations} in frequency space,
transforming functions and operators as $f(\omega) =
\int_{-\infty}^{+\infty}\dt e^{i\omega t}f(t)$ of $f(t)$. Using the
input-output relation \cite{Walls2008} \begin{equation} \label{eq:inout}
a_{\text{out}}(\omega) = \sqrt{2\kappa_E}a(\omega) - a_{\text{in}}(\omega)
\equiv R(\omega)a_{\text{in}}(\omega) , \end{equation} we find the following
solution for the cavity field in terms of the response function of the cavity
\begin{equation} R(\omega) = \frac{2\kappa_E}{\chi_c^{-1}(\omega) +
\sum_{j=1}^N |G_j|^2\chi_j(\omega)} - 1, \end{equation} with the cavity
response $\chi_c(\omega)$ and the mechanical response $\chi_j(\omega)$
functions, given by \begin{equation} \chi_c(\omega) = \frac{1}{\kappa +
i(\Delta -\omega)}, \quad \chi_j(\omega) = \frac{1}{\Gamma_j + i(\omega_j
-\omega)} .  \end{equation}

First, let us briefly discuss the results for driving on the anti-Stokes
sideband, while limiting our discussion to $N=2$ mechanical modes. Without loss
of generality we assume that $G_j$ are real-valued. We furthermore require that
$\Gamma_{j} \ll \kappa$, which is usually the case for typical realizations of
the proposed system \cite{Massel2012, Suh2014}. Also, let us denote the
frequency difference of the mechanical oscillators by $\Delta\omega \equiv
\omega_1 - \omega_m = -(\omega_2 - \omega_m)$. Fig.\ \ref{fig:RedDetuned} shows
the real part of the response function in the vicinity of the resonant
anti-Stokes sideband. For the degenerate case of identical mechanical elements
\cite{Jha2013}, the system divides into a bright superradiant mode (solid
curve) with linewidth $\Gamma_+ = \Gamma +2\Gamma_{\text{opt}}$ and a dark mode
$\Gamma_- = \Gamma -2\Gamma_{\text{opt}}$, which is effectively decoupled from
the cavity. The broadening is proportional to the optomechanical damping rate
$\Gamma_{\text{opt}} = G_j^2 \chi_c$. An analysis of the roots of the
denominator of $R(\omega)$ shows that the formation of a bright and a dark mode
occur only for $\Delta\omega < G_1G_2/\kappa$ and the dark mode asymptotically
decouples from the cavity for $\Delta\omega \rightarrow 0$. We provide a
detailed derivation of this result for the similar case of driving on the
Stokes sideband.

We finally note the connection between the superradiance of the mechanical
oscillators and the superradiance of the atomic system in a cavity
\cite{Haroche1976}. In both cases the system interacts with the cavity photons
of a single common cavity mode which in turn interacts with the vacuum modes of
the outside world -- the bath is made up of all the outside vacuum modes. It is
well known that the single atom decay rate in the cavity is given by the
Purcell formula $g^2/\kappa$ where $g$ is the coupling of the atom to the
cavity. For the mechanical elements, the couling with the cavity gives an
additional decay rate $G^2/\kappa = \Gamma_{\text{opt}}$. In the atomic case,
the decay rate of a single atom in the presence of the other atoms is given by
$Ng^2/\kappa$, whereas in our optomechanical system the decay rate of a single
mechanical oscillator is $NG^2/\kappa$. Thus the similarity between the atomic
and the mechanical case is striking.

As a matter of fact, the whole process can be thought of as a scattering of
phonons of different mechanical oscillators into a common cavity mode of
indistinguishable photons. In this case, we observe superradiance and each
mechanical oscillator is more rapidly damped, \ie emits phonons more rapidly
which are converted into anti-Stokes cavity photons.

\section{Collective Gain} We will now study the system under resonant driving
in the Stokes sideband with $\Delta \approx -\omega_m$. The physical process
here is different -- it leads to the spontaneous generation of a phonon and a
Stokes photon. Thus the phonon field can grow and one can have phonon laser
action.

In what follows, we describe the characteristics and the origin of the
collective behavior resulting in this regime. Using the same approach as
before, we start by dropping non-resonant terms (RWA) in the linearized
optomechanical Hamiltonian \eqref{eq:LinHamiltonian}. This yields the quantum
Langevin equations \begin{subequations} \label{eq:QuLa_blue} \begin{align}
\dot{a} &= -(\kappa + i\Delta) a + i \textstyle{\sum}_{j=1}^NG_jb_j^\dagger +
f_a(t) , \\ \dot{b}_j^\dagger &= -(\Gamma_j-i\omega_j)b_j^\dagger - iG_j^* a +
f^\dagger_{b_j}(t) , \end{align} \end{subequations} with quantum Langevin
forces defined by Eqs.\ \eqref{eq:correlators}. A short calculation yields the
system's response function \begin{equation} \label{eq:BlueResponse} R(\omega) =
\frac{2\kappa_E}{\chi_c^{-1}(\omega) - \sum_{j=1}^N |G_j|^2\chi_j(\omega)} - 1
, \end{equation} with $\chi_c(\omega)$ as above and $\chi_j(\omega) =
1/[\Gamma_j - i(\omega_j +\omega)]$.

For a critical driving power in the Stokes sideband, the system exhibits
self-sustained oscillations \cite{Zhang2014} and becomes unstable. We thus
limit our investigation to the stable driving regime, which can be found by
evaluating the Routh-Hurwitz stability criterion (see, \eg \cite{Hurwitz1964})
for the quantum Langevin equations \eqref{eq:QuLa_blue}. For similar mechanical
oscillators with $\Gamma \approx \Gamma_j$ and $\Delta\omega \approx 0$ the
stability condition evaluates to \begin{equation} \label{eq:Stability}
\textstyle{\sum}_{j=1}^N |G_j|^2 < \Gamma \kappa  .  \end{equation} Under this
condition, the system will also remain stable with finite frequency detuning
$\Delta\omega\neq 0$. For typical structures \cite{Massel2012, Suh2014} we have
$\Gamma \ll \kappa$ and are thus limited to the weak coupling regime, \ie
$G_j^2 \ll (\kappa/2)^2 $, for stable operation. When the condition
\eqref{eq:Stability} does not hold, phonon lasing occurs and Eqs.\
\eqref{eq:QuLa_blue} have to be generalized to include nonlinearities to reach
stable operation \cite{Sargent1974}. With that, the cavity response
$\chi_c(\omega)$ is approximately independent of frequency near the mechanical
resonance frequencies $\omega_j$: \begin{equation}
\label{eq:Approx_Cavity_Response} \chi_c(\omega) \approx \chi_c(-\omega_j) =
\frac{1}{\kappa + i(\Delta + \omega_j)} \approx \frac{1}{\kappa} \, .
\end{equation} The second approximation in Eq.\
\eqref{eq:Approx_Cavity_Response} holds for near-degenerate mechanical
frequencies, \ie $\Delta + \omega_j \ll \kappa$. 

\begin{figure}[htp] \includegraphics[trim=0.5cm 0.7cm 0.4cm 0cm, clip=true,
width=0.44\textwidth]{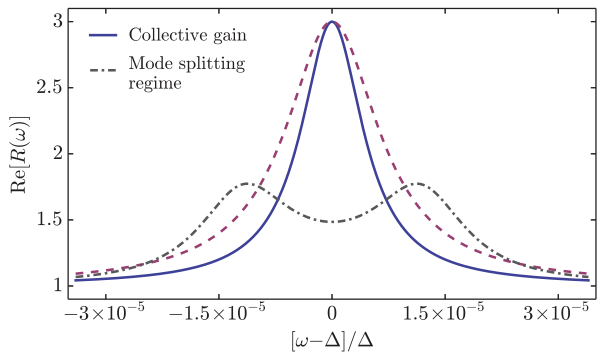} \caption{\label{fig:BlueDetuned}
(Color online) Same as in Fig.\ \ref{fig:RedDetuned} except for driving on the
Stokes sideband with $G=0.5\sqrt{\kappa\Gamma}$. The response function of the
degenerate multimode system (solid curve) shows a decreased linewidth. Here,
the dash-dotted curve corresponds to $\Delta\omega = 1.25\Gamma$.} \end{figure}

In the following, we put special emphasis on the case of $N=2$ mechanical
oscillators and generalize our discussion to more than two modes afterwards.
Fig.\ \ref{fig:BlueDetuned} shows the numerical evaluation of $R(\omega)$ for
this case in the vicinity of the resonance. For the multimode system, the
linewidth of the resonant feature is decreased in comparison to the system of a
single mechanical oscillator, whereas the amplitude is strongly increased
(collective gain). It should be borne in mind that the collective gain, that we
discuss, is in terms of the phonon variables, as we are investigating phonon
laser action. This should be differentiated from the narrowing of the cavity
linewidth which was found in \cite{Xuereb2012}.

\begin{figure}[htp] \includegraphics[trim=0.5cm 0.6cm 0.4cm 0cm, clip=true,
width=0.45\textwidth]{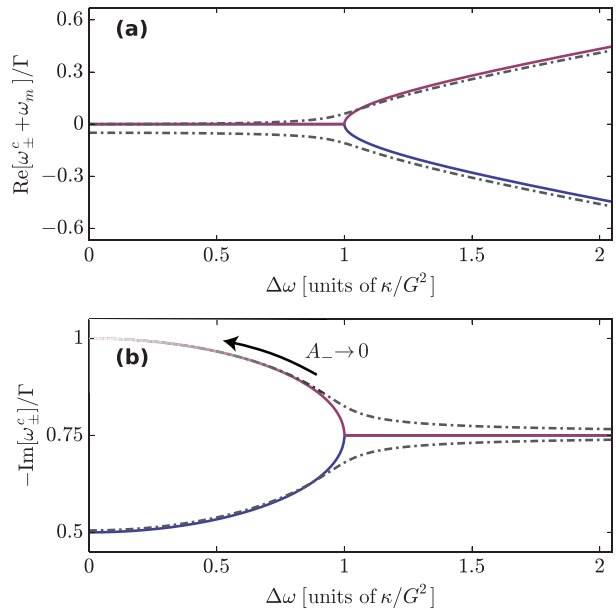} \caption{\label{fig:bifurcation}
(Color online) Real (a) and imaginary (b) part of the denominator roots of
$R(\omega)$ for on-resonance driving on the Stokes sideband. Parameters were
chosen as in Fig.\ \ref{fig:RedDetuned}. Upper red (lower blue) curve
corresponds to the negative (positive) sign in \eqref{eq:Roots_blue}. Dashed
curves were calculated for off-resonant driving with $\delta \equiv \omega_c -
\omega_l + \omega_m = 0.1\kappa$. The bifurcation point vanishes for $\delta
\neq 0$. In (b) the upper curve with amplitude $A_-$ decouples from the cavity
for $\Delta\omega \rightarrow 0$ (see main text for details). Parameters were
chosen as in Fig.\ \ref{fig:BlueDetuned}.} \end{figure}

To get an understanding of the cooperative effects leading to this signature,
we analyze the roots of the denominator of $R(\omega)$. In the approximation of
Eq.\ \eqref{eq:Approx_Cavity_Response} the roots are given by: \begin{equation}
\label{eq:Roots_blue} \omega^c_\pm = \frac{1}{2}\left[\omega^e_1 + \omega^e_2
\pm \sqrt{(\omega^e_1-\omega^e_2)^2 - 4\chi_c^2 G_1^2 G_2^2}\right] ,
\end{equation} with the effective complex frequencies of the mechanical
oscillators $\omega^e_j = -\omega_j - i(\Gamma_j - \Gamma_{\text{opt}})$ and
the optomechanical damping rate $\Gamma_{\text{opt}} = G_j^2 \chi_c$. We can
identify $\chi_c G_1 G_2$ as the effective coupling between the mechanical
modes, mediated by the photonic field. The real and imaginary parts of Eq.\
\eqref{eq:Roots_blue} are shown in Fig.\ \ref{fig:bifurcation}. The square root
term gives rise to two regimes separated by a bifurcation point at
$\Delta\omega = \chi_c G_1 G_2$. For a frequency detuning $\Delta\omega$
smaller than the effective coupling, the effective frequencies of the
mechanical modes become degenerate and form two collective normal modes. This
applies to resonant driving both on the Stokes and on the anti-Stokes sideband.
For off-resonant driving the bifurcation point vanishes; the formation of
collective modes, however, persists. 

A similar case of normal mode splitting is known in the context of strong
coupling between a single mechanical oscillator and the cavity mode
\cite{Aspelmeyer2009}. In our case, however, we are required to work in the
weak coupling regime, where the cavity response function is approximately
independent of frequency in the vicinity of the mechanical resonances. In this
system, the collective modes form due to the effective coupling between the
mechanical oscillators, mediated by the photonic field.

The mode showing collective gain with $\Gamma_+ = \Gamma-2\Gamma_{\text{opt}}$
(lower curve in Fig.\ \ref{fig:bifurcation}) defines the response of the
optomechanical system, as other mode (upper curve) asymptotically decouples
from the cavity for $\Delta\omega \rightarrow 0$. Here, the underlying physical
process can be seen as a scattering from indistinguishable photons in the
cavity field into a single \textit{collective} phonon mode, whereas (in the
case of $N$ mechanical oscillators) the other $N-1$ phononic modes do not
couple to the light field at all. This can also be understood by analyzing
response function $R(\omega)$ in terms of partial fractions \begin{equation}
R(\omega) = 2\kappa_E \left[  \frac{A_+}{\omega - \omega_+^c}
- \frac{A_-}{\omega - \omega_-^c}  -1 \right] -1 .  \end{equation} This
  analysis shows that $A_- \rightarrow 0$ for $\Delta\omega \rightarrow 0$.
This also applies to the anti-Stokes sideband, where only the superradiant mode
with $\Gamma_+ = \Gamma + 2\Gamma_{\text{opt}}$ contributes to the response
function. 

In a somewhat different picture this effect can be understood in terms of the
structure of the response function. $R(\omega)$ of the fully degenerate system
with identical mechanical oscillators is equal to the response function of a
system with a \emph{single} oscillator in a cavity with an effective coupling
rate of $G' = \sqrt{2}G$. This applies both to the Stokes and the anti-Stokes
sideband and can readily be generalized for a system of $N$ oscillators. Here,
the effective coupling rate reads $G' = \sqrt{N}G$. In the same way,
generalization of the collective linewidth yields $\Gamma_+ =  \Gamma \pm N
\Gamma_{\text{opt}}$, whereas the plus (minus) sign holds for the anti-Stokes
(Stokes) sideband. This implies that the threshold power for phonon laser
action would drop by a factor of $N$ \footnote{H. Jing et al., Phys. Rev. Lett.
\textbf{113}, 053604 (2014) have noticed the possibility of ultrathreshold
phonon lasing using PT symmetric physics. Our work differs from theirs as we do
not use PT symmetry.}.

\begin{figure}[htp] \vspace{1em}\includegraphics[trim=0.5cm 0.7cm 0.4cm 0cm,
clip=true, width=0.44\textwidth]{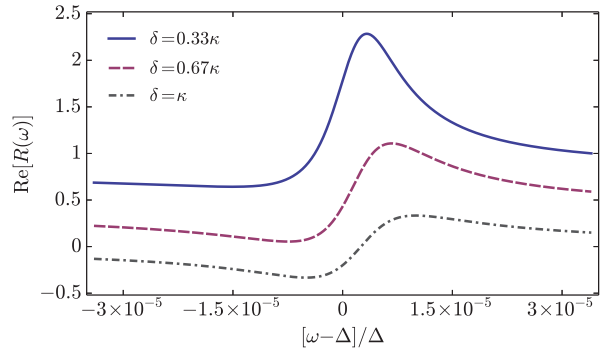}
\caption{\label{fig:detuning} (Color online) Real part of response function for
off-resonant driving with detuning $\delta \equiv \omega_c - \omega_l +
\omega_m$ in the vicinity of the Stokes sideband. Parameters were chosen as in
Fig.\ \ref{fig:BlueDetuned}.} \end{figure}

For off-resonant driving the collective behavior persits, the response function
however gradually changes from an absorptive to a dispersive structure, as
shown in Fig.\ \ref{fig:detuning}. The resonant feature considerably broadens
and decreases in amplitude. Qualitatively, this behavior occurs for any number
of oscillators and is not distinctive for the multimode system. In the here
discussed case of frequency degeneracy, the system of N mechanical oscillators
with coupling rates $G$ is equal to a system of a single oscillator with a
coupling rate of $\sqrt{N}G$.

\section{Conclusions} In summary, we have derived and analyzed the emergence of
superradiance and collective gain in optomechanical multimode systems. We have
discussed the necessary conditions for these collective effects to occur and
outlined the experimental feasibility. In the light of recent progress in
on-circuit implementations of optomechanics \cite{Massel2012, Suh2014}, the
experimental realization of the proposed system should be within reach. We
showed that the system can easily be switched from a superradiant state,
allowing enhanced cooling of a single vibrational mode, to a collective gain
regime where phonon laser action \cite{Painter2010} with a single collective
phononic mode is potentially possible.

\begin{acknowledgments} We would like to thank Junho Suh for providing us with
information about their multimode optomechanical device. T.\ K.\ gratefully
acknowledges support and advice from Joachim von Zanthier, the German National
Academic Foundation and the hospitality of the Oklahoma State University, while
this work was done.  \end{acknowledgments}

%
\end{document}